\documentclass[10pt,towcolumn]{IEEEtran}
\usepackage{setspace}

\usepackage{cite}

\usepackage[pdftex]{graphicx}

\usepackage{url}

\begin{document}

\title{Measurement of the Optical Turbulence Produced by a Multirotor Unmanned Aerial Vehicle}

\author{Ra\'ul Rodr\'iguez Garc\'ia,
        Luis Carlos Alvarez Nu\~nez,
        and~Salvador Cuevas
\thanks{This work was done with the support of Consejo Nacional de Ciencia y Tecnologia (CONACYT) student grant 373802 and the PAPIIT UNAM program grant IT101116.}        
       
\thanks{R. Rodr\'iguez Garc\'ia is with the Department
of Astronomical Instrumentation of Institute of Astronomy, National Autonomous University of Mexico, Mexico, e-mail: rrodriguez@astro.unam.mx.}
\thanks{L. C. Alvarez Nu\~nez work at Institute of Astronomy, National Autonomous University of Mexico, Mexico, e-mail: lalvarez@astro.unam.mx.}
\thanks{S. Cuevas work at Institute of Astronomy, National Autonomous University of Mexico, Mexico, e-mail: chavoc@astro.unam.mx.}}

\maketitle

\begin{abstract}
At present, new approaches for the use of Multirotor Unmanned Aerial Vehicle or multirotor drones in high precision optical applications are rising. However, the optical turbulence effects generated by multirotor drones are not entirely understood. These optical effects can reduce the performance of the optical instruments that they transport. This paper presents measurements of the wavefront deformation generated by the temperature fluctuations and the airflow of a drone's propulsion system. To do so, we used a single arm of a DJI S800 EVO Hexacopter (professional drone) and measured its operating temperature with a commercial infrared camera. The resulting temperature variation, between a switched-off propulsion system at room temperature and one running at its maximum performance, was 34.2° C. Later, we performed two different interferometric tests, Takeda's method, and the phase-shifting technique, using a ZYGO interferometer.  These tests show that the total deformation over an incident wavefront to the propeller airflow is lower than 0.074 \boldmath$\lambda$ PV and 0.007 \boldmath$\lambda$ RMS (HeNe laser, $\lambda$=633nm). We conclude that the optical turbulence produced by a drone propulsion system is negligible.
\end{abstract}


%

\section{Introduction}

The development of multirotor drones has been astounding, and today we can find them in a great variety of scientific applications \cite{Kyrkou2019}. The crucial point of this expansion has been the implementation of more robust and precise flight controllers, as well as the improvement of the battery technologies. These have increased drone's maneuverability (even automatically) and flight time, thereby easing their professional use.

Flight performance of multirotor drones has become more stable and accurate \cite{Benassi2017}. This development has inspired new applications involving the use of high precision optics, such as in astronomical instrumentation.  Here,  multirotor drones need to carry a light source that will use as a reference source for astronomical telescopes’ applications. An example of these applications can be: the maintenance of telescopes, optical characterization or adaptive optics \cite{Biondi2016} \cite{Basden2018} \cite{Rodriguez-Garcia2018}.

In optical instrumentation, there are several reasons why an optical system cannot reach its ideal performance. Sometimes inhere in their design and manufacturing parameters, and other times are resulting from external factors such as vibrations, temperature variations of their optomechanical components, and optical turbulence. The latter is produced by random variations in the refractive index of air due to changes in its density or temperature. These variations result in lower quality of the images obtained by the optical instruments. These variations are common and occur as a natural phenomenon in the atmosphere. This event is called atmospheric turbulence or seeing.

Seeing's effects depend on the interaction of air layers of different temperatures. This interaction produces optical turbulence in the form of randomly moving cells of air with different sizes and refraction indexes. When an incident wavefront refracts through those cells, it distorts. Then, this perturbed wavefront arrives on the input pupil of an optical system and blurs the formed image at the instrument focal plane. The strength of blur depends on the relative size of the cells, the wavelength, and the pupil diameter.

If the size of atmospheric cells is larger than the input pupil diameter, a perfect optical system will produce Point Spread Function (PSF) images determined by the diffraction limit of the pupil. When the size of atmospheric cells is smaller than the pupil diameter, such that the number of encircled cells is bigger than 3-4, the PSF energy will be transferred from the central core to the diffraction rings at a rate of change determined by the velocity displacement of cells over the pupil. 

This reasoning can be applied to the air flux produced by a drone propeller and its motor (propulsion system): The mixed layers circulating near a hot motor and the air layers propelled around it could generate optical turbulence. If the temperature difference among the mixed layers is large enough, the size of the cells could affect the wavefront transmitted from a light source placed on the drone. 

To determine if a drone propulsion system produces optical turbulence, a first attempt would be numerical modeling of the air layers around the motor. This analysis would require knowing in advance the differences in temperature of the motor and the surrounding air, as well as the parameters of propeller and characteristics of air. We could use a multiphysics numerical code based on finite element calculations. Nevertheless, the best approach is by measuring the optical turbulence produced by a drone propeller and its motor. That means to measure directly the wavefront distortions using instruments and techniques with enough optical sensitivity. 

The purpose of this paper is to determine the atmospheric turbulence produced by a drone's propulsion system. To this end, we conducted three different optical tests: a Schlieren imaging test, and two interferometry tests (using the Takeda method and the phase-shifting technique). Additionally, we measured the temperature of the motor and the surrounding air using a thermal imaging infrared camera. These type of cameras have an appropriate image resolution (320 X 240 pixels) and temperature sensitivity of around 0.1°C per pixel.

Section \ref{sec:tempmessure} shows the measurements of the increase of temperature of the motor, running at its maximum power, using an infrared camera. Section \ref{sec:Schlieren} addresses the analysis of the distribution of turbulent airflow with Schlieren imaging test. Section \ref{sec:Wavemeasure} presents the development of an experiment using two different interferometric tests. This was because the propulsion system generated many vibrations and we wanted to be sure about the obtained results from this experiment. Finally, we give a summary and our conclusions in Section \ref{sec:conclu}.

\section{Motor temperature variation}
\label{sec:tempmessure}

\begin{figure}[ht]
\centering
\begin{tabular}{cc}
   \includegraphics[width=\linewidth]{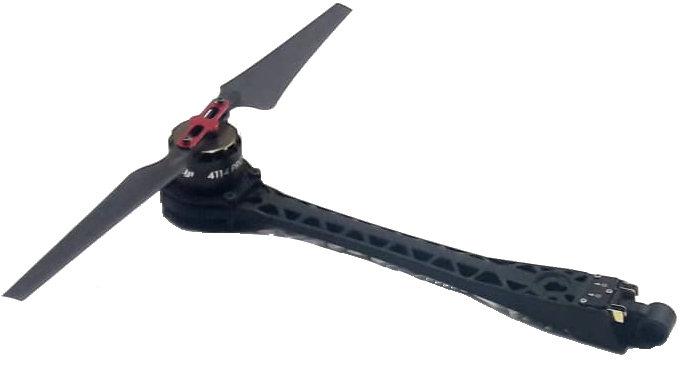}
 \end{tabular}
  \caption{DJI S800 EVO propulsion system.}
  \label{fig:motor}
\end{figure}

To evaluate the change in temperature of drone's propulsion system, we used a single arm of a DJI S800 EVO drone (see figure \ref{fig:motor}). Its features are described below:

\begin{itemize}
\item Motor.- Model: DJI-4114 400 Kv\footnote{Kv is a parameter used by motor manufacturers to characterize  the electro-mechanical performance of their motors.}, max power: 500 W.
\item Electronic Speed Controller (ESC).- Max Current: 40 A, operating voltage: 22.2 V.
\item Foldable propeller.- Engineering plastic, size: 15 X 5.2 in.
\end{itemize}

As mentioned above, it is necessary to have temperature differences within an air flow to induce a distortion on an incident wavefront. We determined the change of temperature of the drone’s propulsion system using a FLIR E4 infrared camera. 

Usually, the motors of a multirotor drone receive operating signals directly from the drone’s flight controller. For the tests, we needed to run one motor independently, so we developed an electronic control board to adjust the speed of the propulsion system as required. With this control board, we also monitored the motor electrical parameters (voltage, current, and power)  during the experiment.

Firstly, we measured the drone's propulsion system at room temperature when the motor was switched off. Then, we made another measurement after five minutes of motor operation at maximum speed. 

\begin{figure}[ht]
\centering
\begin{tabular}{cc}
  \includegraphics[width=42mm]{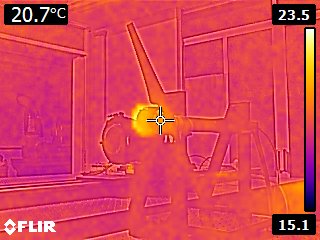}
  \includegraphics[width=42mm]{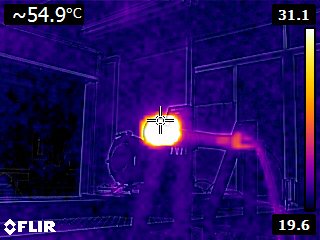}
 \end{tabular}
  \caption{Difference in the temperature of the motor when it was switched-off (left) and after five minutes of operation at its maximum speed (right).}
  \label{fig:temp}
\end{figure}

Figure \ref{fig:temp} shows two thermal images taken with the infrared camera. The measurement highlighted on the upper left side of both images represents the temperature in the central spot. It is worth mentioning that the laboratory has a controlled temperature of 20 °C. This condition ensures that the surrounding air has the same conditions.

It is not unusual to find this, both in addition to the motor and in a separate form, electronic devices used to control and power  the propulsion system are other sources of heat. However, in the case of our experiment, these elements are included in the base of the motor (see DJI S-800 User's manual \cite{DJI2014}), so the obtained measurements  incorporate  the total amount of the generated heat. The difference in temperature from the comparison of both images is 34.2 °C (see figure \ref{fig:temp}).

Besides, we can compare the obtained value with the performance of a propulsion system with similar specs on the T-Motors company website. We found the datasheet of the motor model MN4014 400 Kv with a carbon fiber propeller 15X5. This motor reaches an operating temperature of 46 °C after ten minutes of use at its maximum power  \cite{T-motor}. This temperature value is consistent with our measurements.

\section{Distribution of turbulent flow}
\label{sec:Schlieren}

To better understand the distribution of the heated air flux produced by the propulsion system, we performed a Schlieren test. This kind of test has been widely used to study air flux related problems \cite{Settles1995}. For this test, we used the largest mirror available in our laboratory (60 cm), since the area covered by the rotating propeller is 40 cm in diameter. In this qualitative test, we ran the motor to its maximum speed for about five minutes and then we reduced this speed by half to perceive the optical effects of turbulence.

\subsection{Schlieren Test Setup}

\begin{figure}[htbp]
\centering
\fbox{\includegraphics[width=75mm]{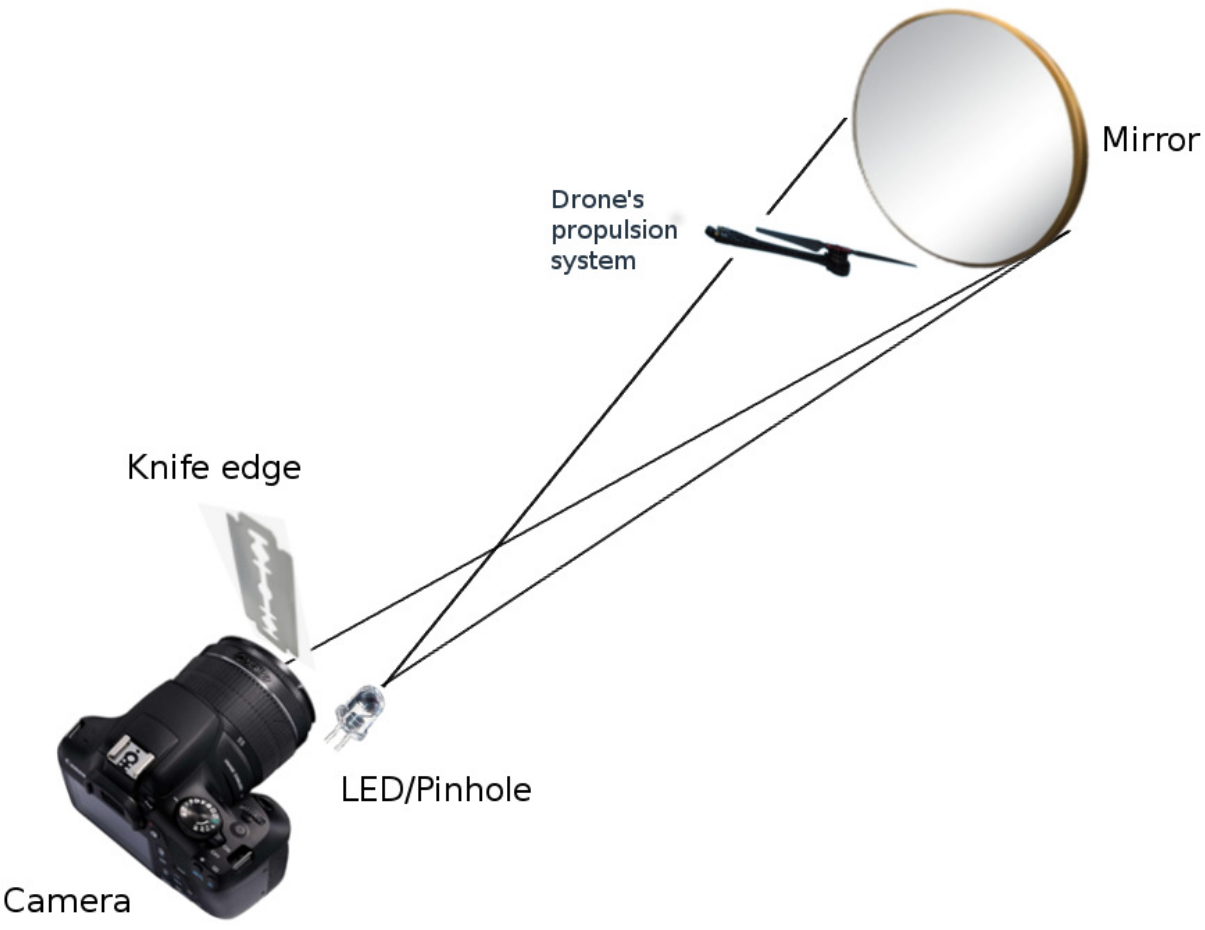}}
\caption{Setup of the Schlieren test performed to get information on the distribution of the optical turbulent flow.}
\label{fig:Schlieren}
\end{figure}

We implemented the Schlieren test with a double pass coincident setup (see figure \ref{fig:Schlieren}) with a spherical mirror 60 cm of diameter and 4.4 m of focal distance. The drone propulsion system was in front of the mirror at a distance of 50 cm. At this distance,  we positioned the propulsion system above and below to the mirror image formed on the camera's CCD (Schlieren test area). This configuration would allow the air flux to cross through the total test area. The light source was a circular pinhole of 1 mm in diameter illuminated by a white LED (1 W high power).  We acquired the images using a Canon T3i camera with a 22,3 X 14,9 mm CMOS detector and a 50 mm objective lens focused on the motor.

\subsection{Results of the Schlieren Test}

First, we placed the propulsion system above the test area. In this configuration, the downward air produced by the propellers crossed the entire mirror. Nevertheless, we could not detect any variation (turbulence) in the Schlieren image. 

\begin{figure}[htbp]
\centering
\begin{tabular}{cc}
  \includegraphics[width=42mm]{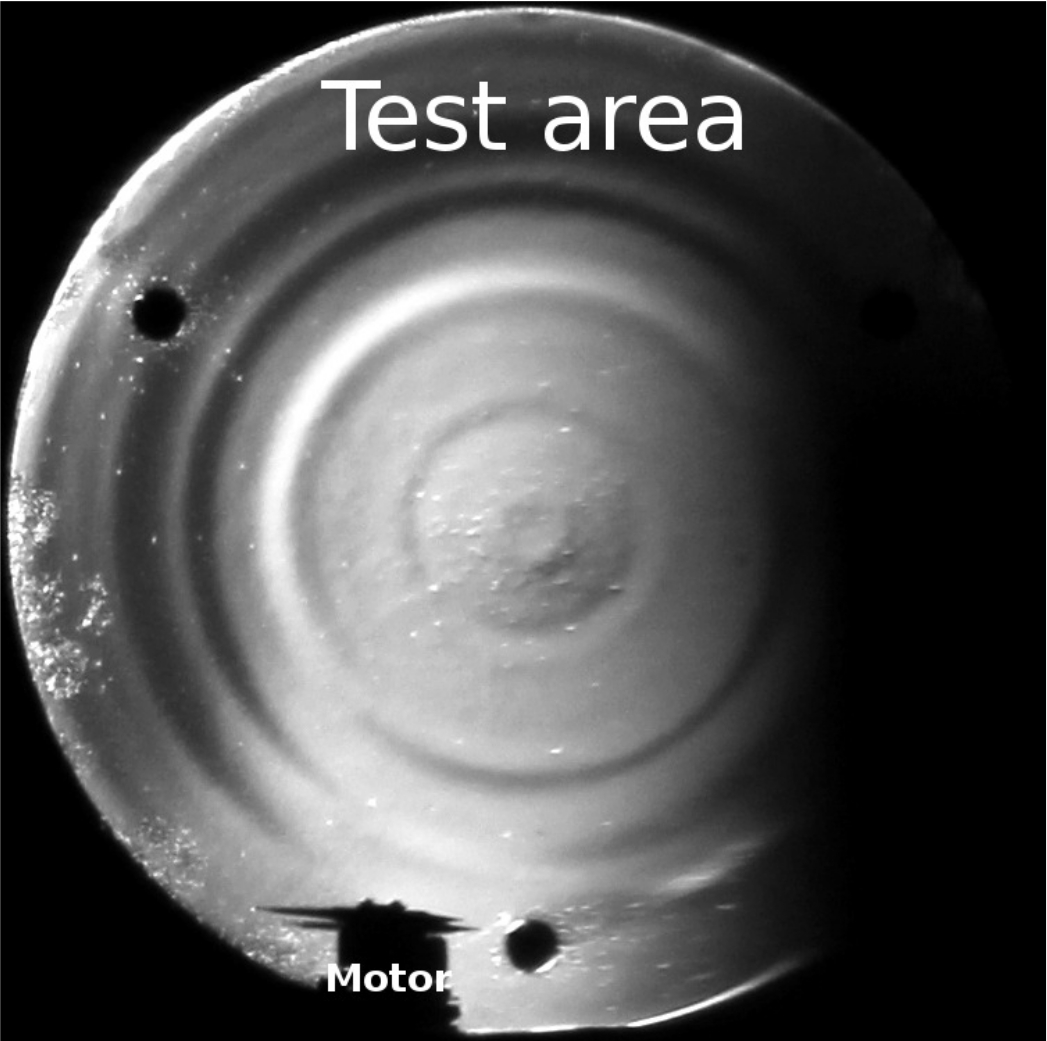}
  \includegraphics[width=42mm]{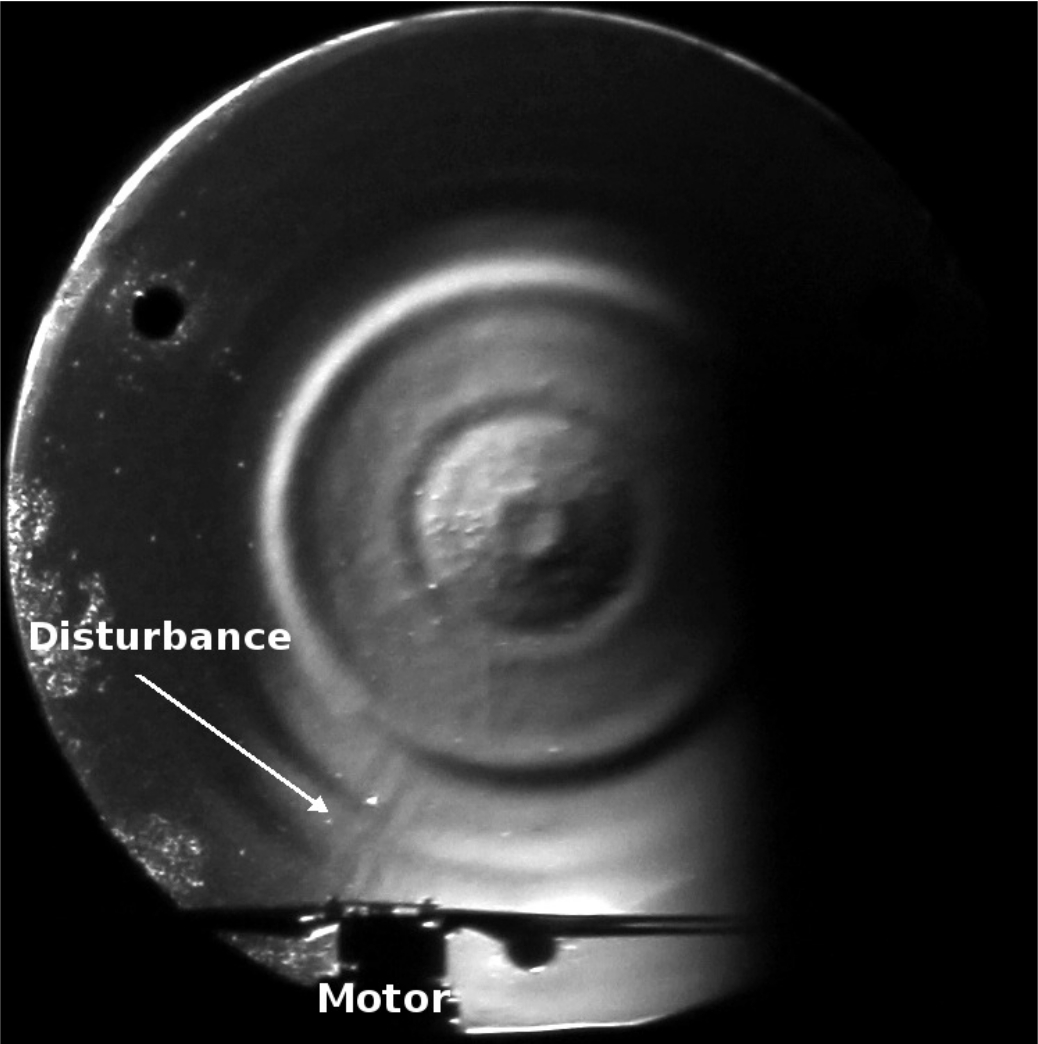}
   \end{tabular}
\caption{Schlieren test with an operating motor (left) and the motor turned  off (right). The right image shows the turbulence produced by the heat of the stopped motor after running. The circular patterns are due to defects in the polishing of the mirror.}
\label{fig:Schlieren1}
\end{figure}

Next, we placed the propulsion system below the test area, under the assumption that the heated air was probably going to move upwards. Again, we could not detect any turbulence after running the motor for five minutes.  Nonetheless, only when we switched off the motor, the camera registered fluctuations produced by the ascending heat (see figure \ref{fig:Schlieren1}).

It was not possible to see optical turbulence using the Schlieren test. Therefore, it was not possible to detect its distribution. These preliminary results led us to infer that the turbulence must be close to the motor. Hence, for the interferometric tests, we decided to place the propulsion system as close as possible to the light beam (see figure \ref{fig:exp-setup}).

\section{Wavefront measurements}
\label{sec:Wavemeasure}

This section describes the experiment that we implemented  to estimate the optical turbulence produced by the drone's propulsion system. We proposed to use interferometric methods since they are the best ways to detect and quantitatively measure the smallest variations (smaller than $\lambda/10$) of wavefront respect to a reference surface.

To measure wavefront distortions, we did two interferometric tests using a 6 inches Fizeau interferometer (ZYGO interferometer) with a high-performance transmission flat ($\lambda$/20). It should be mentioned that the interferometer of our  optical laboratory is certified by the National Institute of Standards and Technology (NIST). The first interferometric test was the Fourier interferometric fringe pattern analysis, also known in optics as the Takeda's method  \cite{Takeda1982}. The second test was using the phase-shifting technique made directly by ZYGO software. We used MATLAB to process and display the phase maps obtained in both tests. The root mean square (RMS)  and Peak to Valley (PV) units are in $\lambda$ of a HeNe laser ($\lambda$=633 nm). 

In both tests, the motor ran to its maximum speed for five minutes. Its electrical parameters were: voltage of 21.3 V, current of 19.79 A and 418 W of power (the maximum specified operating power is 500 W). Next, we reduced the motor speed by half (20.9 V, 5.4 A, and 113 W). Then, we made eight measurements during the next five minutes. This procedure simulates the flight of a drone until it reaches a required operating height, and then maintaining a static flight (hover mode) to perform some task.

\subsection{Experiment Setup}

\begin{figure}[ht]
\centering
\fbox{\includegraphics[width=\linewidth]{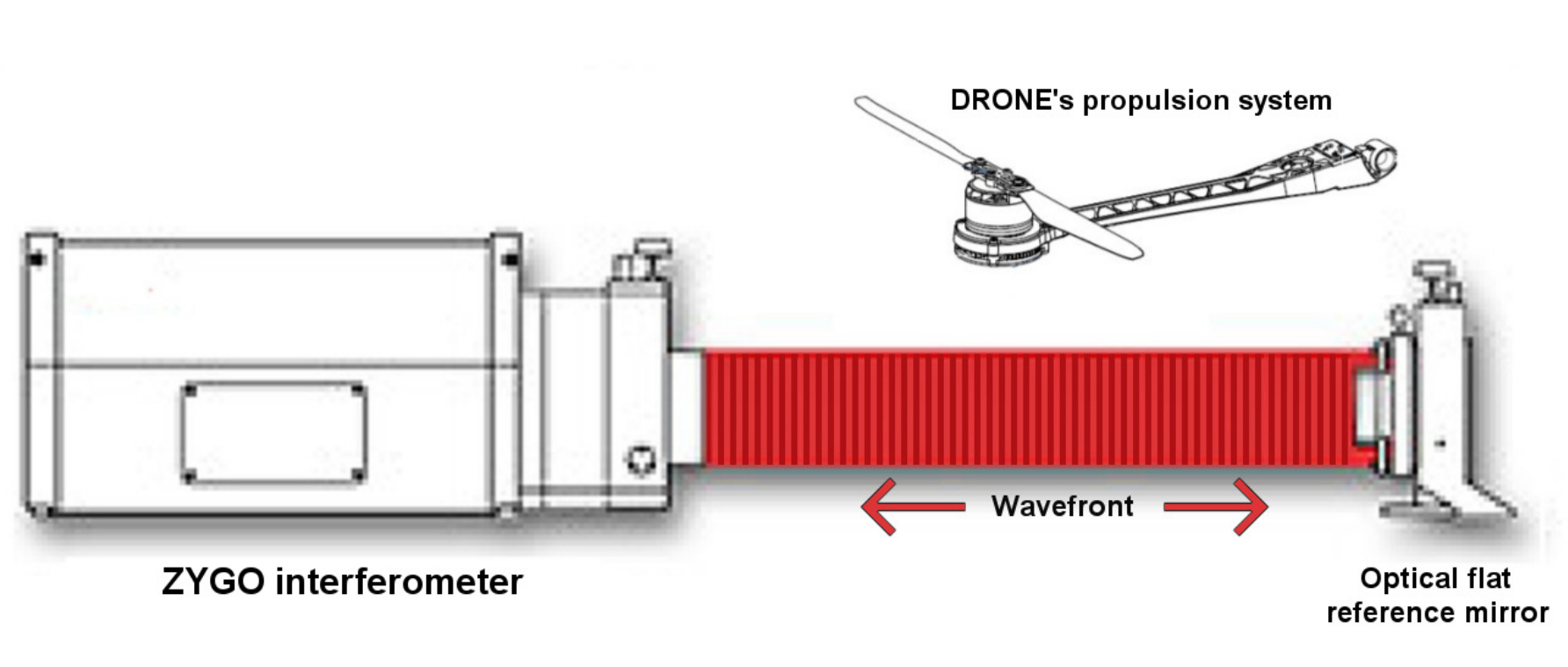}}
\caption{Interferometric experiment layout.}
\label{fig:exp-setup}
\end{figure}

In the experiment setup, an external 4 inches flat reference mirror ($\lambda/12$ PV) reflected the light beam from the interferometer by the same optical path. This arrangement allows us to have a reference wavefront. Figure  \ref{fig:exp-setup} shows a layout of the performed experiment.

\begin{figure}[ht]
\centering
\fbox{\includegraphics[width=\linewidth]{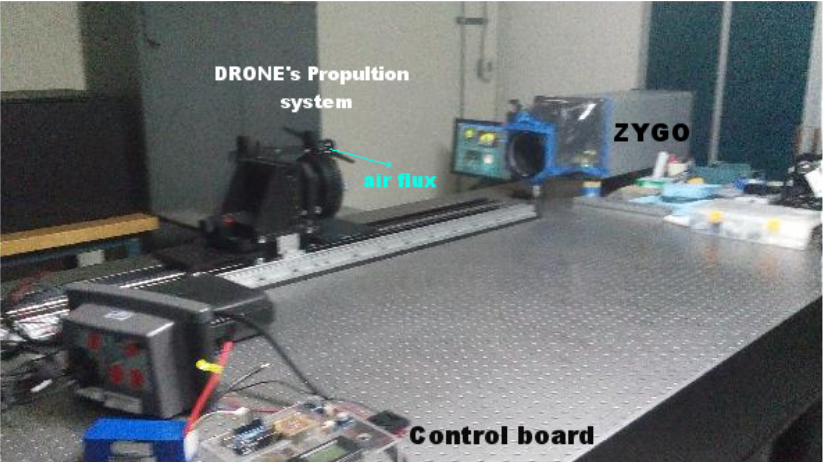}}
\caption{Experiment Setup. The propulsion system was isolated from the optical table, and we operated it by the control board. The propeller was placed laterally to the light beam of the interferometer, directing the hot air flow into the laser beam.}
\label{fig:experiment}
\end{figure}

With this experimental configuration, any small temperature variation distorts the wavefront in a double pass. This configuration doubles the sensitivity of the experiment. We placed the drone propulsion system at a distance of 35 cm from the light beam with the rotation axis of the propeller perpendicular to the optical beam direction. The purpose of this was that the heated air crosses the beam of light (see figure \ref{fig:experiment}). It should remember that the controlled temperature of the laboratory was of 20 °C.

\subsection{Analysis with the Takeda's Method}

The first interferometric test was Takeda's method due to the fact that the normal operation of the drone's propulsion system produces vibrations. The Takeda's method is very effective in the presence of small mechanical vibrations as it uses only one image from the interference fringe pattern or interferogram to obtain the phase-map information. The interferogram freezes the instantaneous characteristics of the wavefront, including the turbulence effects at this moment.

\begin{figure}[htbp]
\centering
\fbox{\includegraphics[width=50mm]{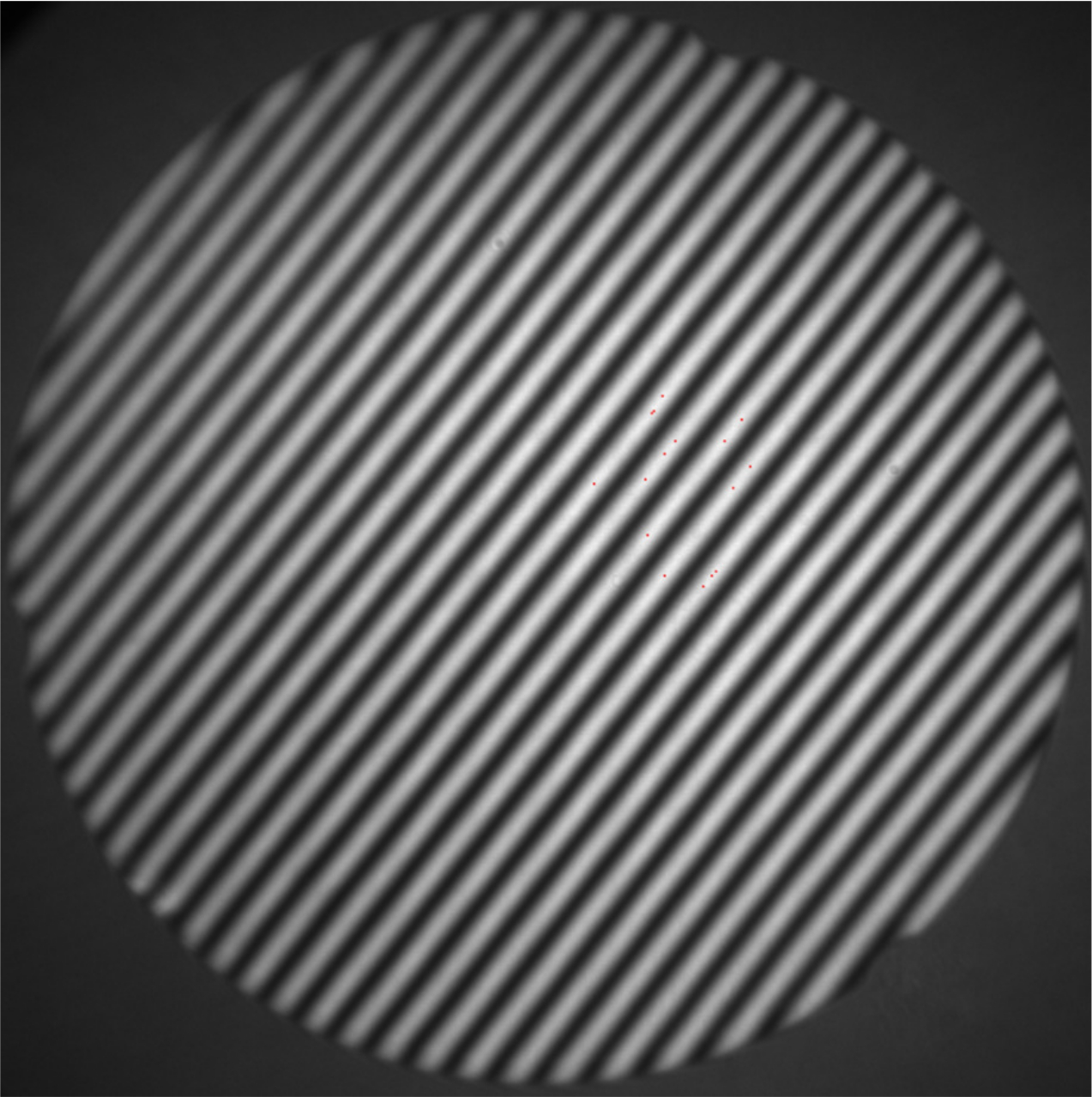}}
\caption{Interferogram of the reference flat mirror with a small tilt.}
\label{fig:takedafringe}
\end{figure}

It is important to have a considerable number of fringes so that the algorithm can work correctly \cite{Macy1983}. We modified the number of obtained fringes by small adjustments in the tilt of the external flat reference mirror. Figure \ref{fig:takedafringe}  shows an example of a fringe pattern obtained for Takeda's method test.

We implemented Takeda's method in MATLAB, to get the phase-maps from the interferograms. It is worth mentioning that we needed to calibrate our software to get the actual values of Peak to Valley and RMS from the phase-maps' measurements.

We did the calibration of our software by comparing one of its post-processed phase-maps with one obtained by a standard measurement of ZYGO interferometer (phase-shifting). Both of them were from the flat reference mirror under the same conditions.  In the calibration process, we performed the measurements without the influence of the propulsion system and in a controlled temperature environment. 

\begin{figure}[ht]
\centering
\fbox{\includegraphics[width=\linewidth]{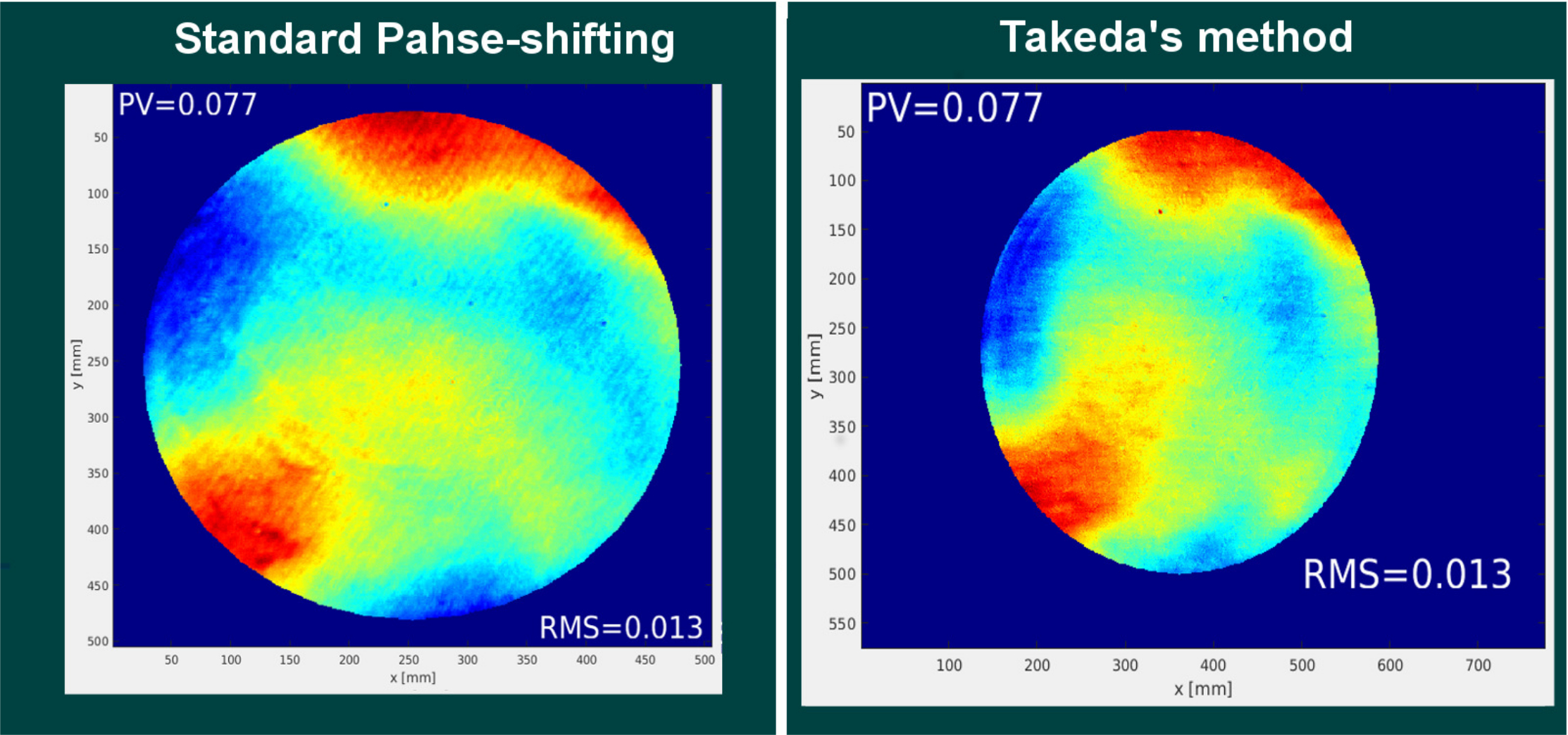}}
\caption{Phase-maps of the flat reference mirror used to calibrate the software. The image at the left shows a standard phase-shifting measurement, and the image at the right shows a measurement with Takeda's method software after adjusting the values of PV and RMS.}
\label{fig:takedacalib}
\end{figure}

Figure \ref{fig:takedacalib} shows the phase-map of the flat reference mirror obtained by the two previously described measurements over the same test area. Here, the wavefront errors after the calibration were the same, 0.077 $\lambda$ PV and 0.013 $\lambda$ RMS.

For each measurement, we subtracted the phase-map obtained without the impact of the propulsion system (the calibration phase-map) to isolate the turbulence effects in the tests. This subtracted phase-map includes the errors of the flat reference mirror. Therefore, we can consider it as the instrumental error.

We ran the propulsion system for five minutes at its maximum speed an then we reduced this speed by half. Then, we took eight interferograms over the next five minutes to be post-processed. 

\begin{figure*}[htbp]
\centering
\fbox{\includegraphics[width=0.95\textwidth]{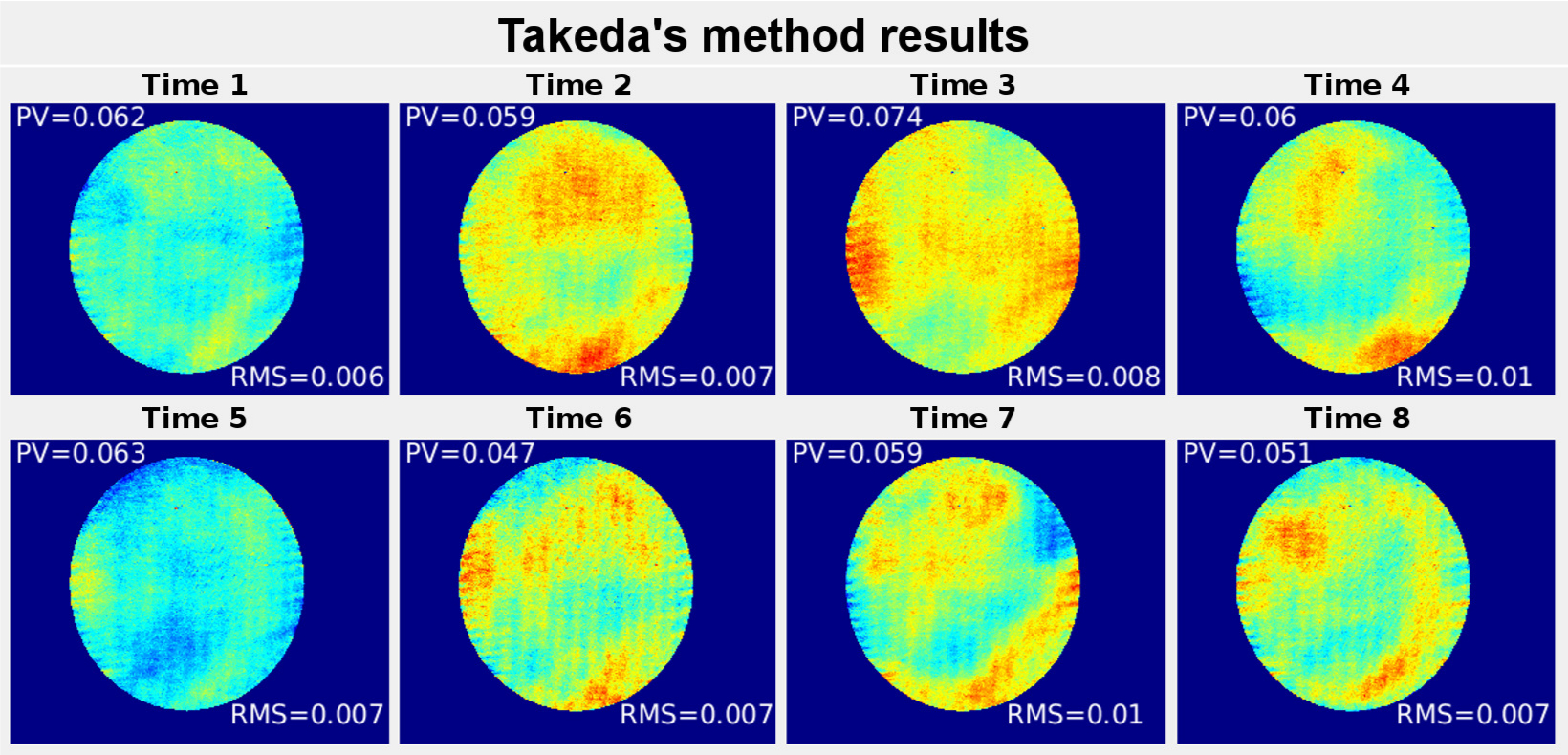}}
\caption{Experiment's phase-maps obtained with the Takeda's method software.}
\label{fig:takedares}
\end{figure*}

Figure \ref{fig:takedares} shows the results achieved by the implemented software. The min and max values of RMS were 0.006 and 0.01 $\lambda$,  respectively. The mean PV of all the events was 0.053 $\lambda$ ($\lambda/19$) and the mean RMS is 0.007 $\lambda$, equivalent to a Strehl Ratio of 0.998 (using Marechal's formula).

\subsection{Analysis with the Phase-Shifting Technique}  

\begin{figure}[htbp]
\centering
\fbox{\includegraphics[width=50mm]{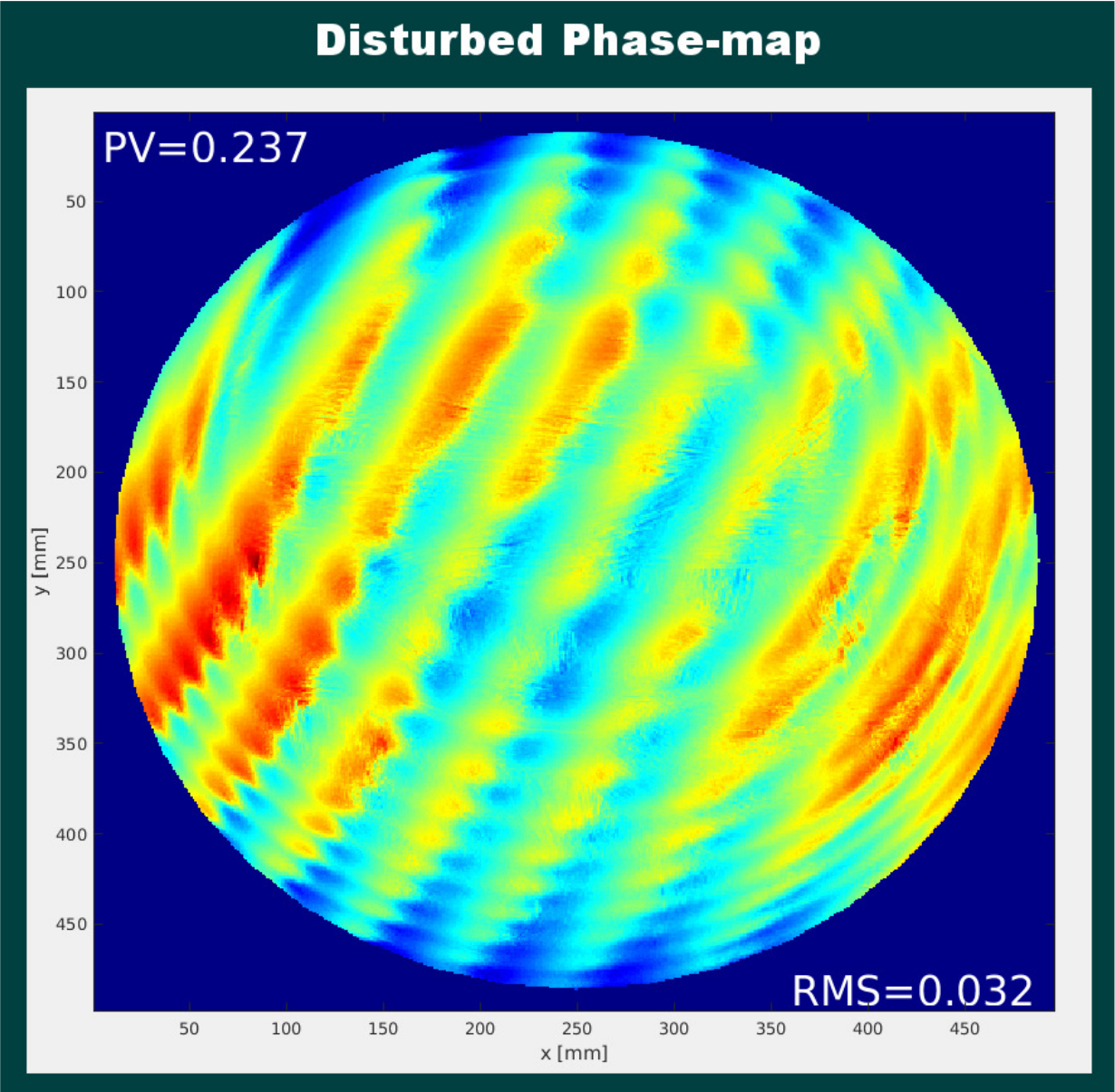}}
\caption{Phase-map showing the disturbance produced by the vibration of the optical table.}
\label{fig:vibration}
\end{figure}

Complementary to this work, we verified the results obtained by Takeda's method. To do this, we performed direct interferometric measurements with the ZYGO instrument. However, even though the motor was isolated from the optical table, the propulsion system produced vibrations by the ejected air from the propeller to the optical table. We made standard phase-shifting measurements, but the resulted phase-maps were distorted. Figure \ref{fig:vibration} shows the distortion of the obtained phase-map. Here, the wavefront errors were of 0.237 $\lambda$ PV and 0.032 $\lambda$ RMS.

In order to eliminate the effects of vibrations, we modified the acquisition time parameter of the interferometer camera from 2000 $\mu$s (default value) to 5 $\mu$s. However, when conducting various measurements on the reference flat mirror, we found variations in the resulting phase-maps, even without the impact of the propulsion system. We attributed these variations to the reduction of the signal-to-noise ratio resulting from the modification of the exposure time.

Similarly to in the Takeda's test, we isolated the effects of turbulence subtracting the instrumental error from each of the measurements. In this case, the instrumental error was the average of ten short exposure phase-maps made with the ZYGO interferometer software. Then this error was subtracted from each measurement using the ZYGO software too. 

\begin{figure*}[htbp]
\centering
\fbox{\includegraphics[width=0.95\textwidth]{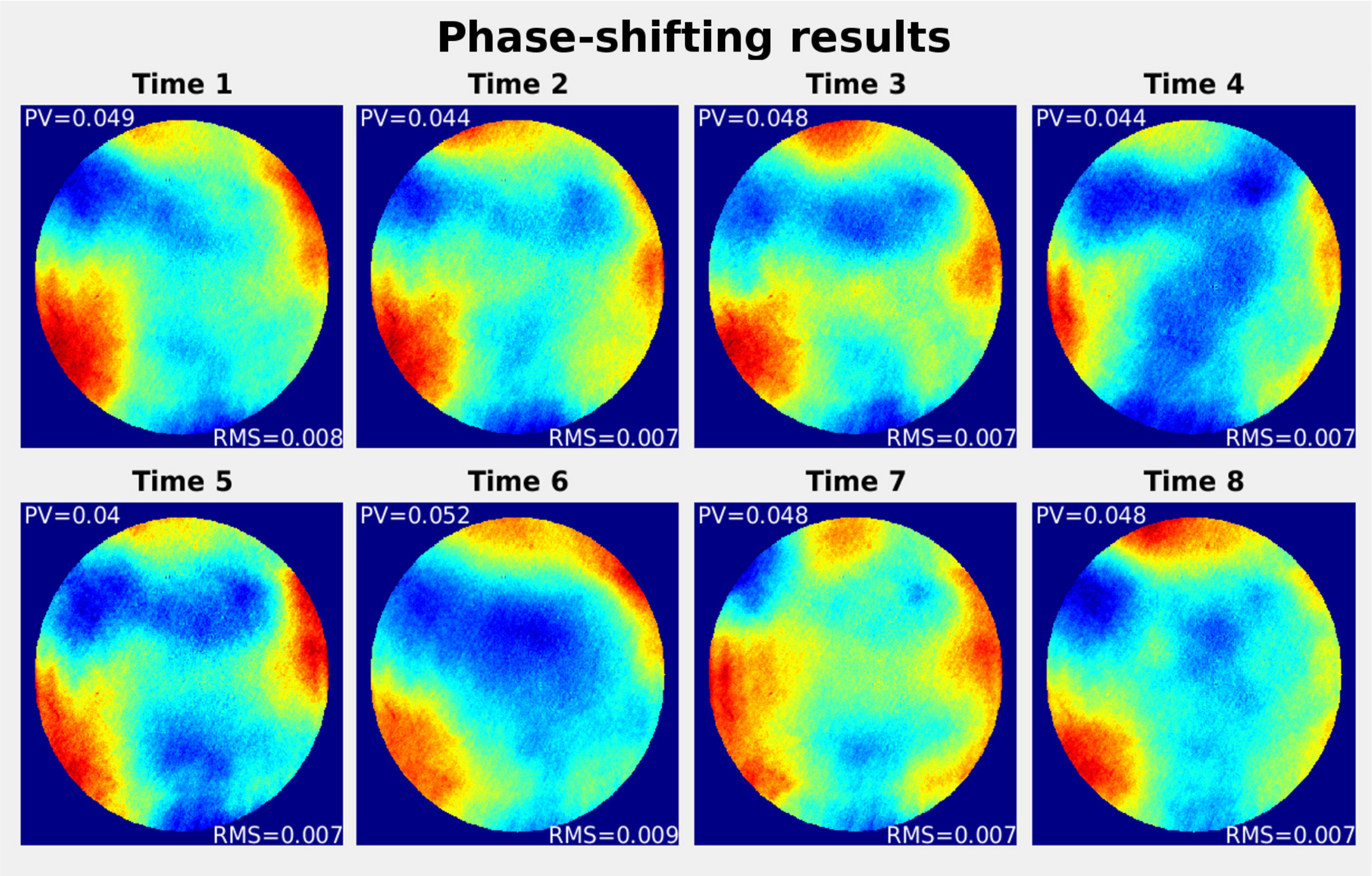}}
\caption{Experiment's phase-maps obtained with the phase-shifting technique with short exposure time.}
\label{fig:res}
\end{figure*}

In figure \ref{fig:res} we show the results of the phase-shifting technique measurements. In this case, we have a mean PV of 0.046 $\lambda$ ($\lambda/21$) and min/max RMS values of 0.007 and 0.009 $\lambda$, respectively, with a mean RMS of 0.007 $\lambda$, equivalent to a Strehl Ratio of 0.998.

\newpage
\section{Summary and Conclusion}
\label{sec:conclu}

We measured the temperature of a drone propulsion system (motor, propeller, and electronics) using an infrared camera. We obtained a gradient of 34.2 °C  after running the propulsion system at its maximum speed for 5 minutes.

We conducted a Schlieren test to determine the distribution of turbulence flow. In this test, we did not see the optical turbulence. However, we observed optical disturbances produced by the heat of the motor when we switched-off the propulsion system after running the test.

We also conducted two interferometric tests using a ZYGO Interferometer: one performed with the post-processing of interferograms using a  Takeda's method software another using the phase shifting technique with a modified acquisition time.   

The results of all the test show random variations between each phase-map. The wavefront errors are below the instrumental error. The values of RMS and PV obtained in the fast phase-shifting test, are comparable to those obtained with the Takeda's test. 

As a result of our experiment, we can affirm that the propulsion system does not produce significant optical turbulence. Therefore, drones can be used in high-precision optical applications.

\section*{Acknowledgment}

The authors are thankful for the facilities offered by the institutions Instituto de Astronomia, UNAM, and SEREPSA, UNAM. We also  would like to thank Dr. Jorge Fuentes for his detailed review which improved our manuscript.

\begin{IEEEbiography}[{\includegraphics[width=1in,height=1.25in,clip,keepaspectratio]{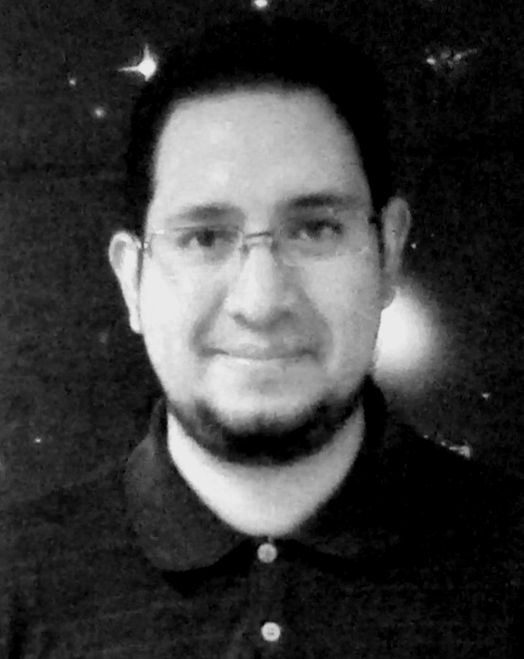}}]{Ra\'ul Rodr\'iguez Garc\'ia} received the B.E.(Mechatronics) and M.S.E.E.(Instrumentation) from the National Autonomous University of Mexico (UNAM). He has developed optical instruments applied to astronomical instrumentation at the Institute of Astronomy UNAM. He worked in the manufacture of the optical components of the FRIDA instrument, a collaboration project between the IA-UNAM and the 
Institute of Astrophysics of the Canary Islands (IAC). He is currently studying new applications of Drones for astronomical instrumentation.
\end{IEEEbiography}

\begin{IEEEbiography}[{\includegraphics[width=1in,height=1.25in,clip,keepaspectratio]{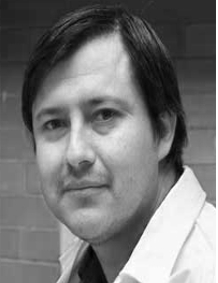}}]{Luis C. Alvarez Nu\~nez} studied electronics engineering at the Technological Institute of Durango, Mexico developing a He-Ne laser; Later he completed his Masters and PhD studies at the Optics Research Center in Guanajuato, Mexico, in the area of manufacturing and optical testing. He participated in the first ISCAI2009 Advanced Instrumentation school in the Canary Islands, Spain, working on the M5 mirror control system for the E-ELT at the company NTE-SENER, in Barcelona, Spain. He is currently working in the development of optical instruments at the Institute of Astronomy of the National Autonomous University of Mexico.
\end{IEEEbiography}

\begin{IEEEbiography}[{\includegraphics[width=1in,height=1.25in,clip,keepaspectratio]{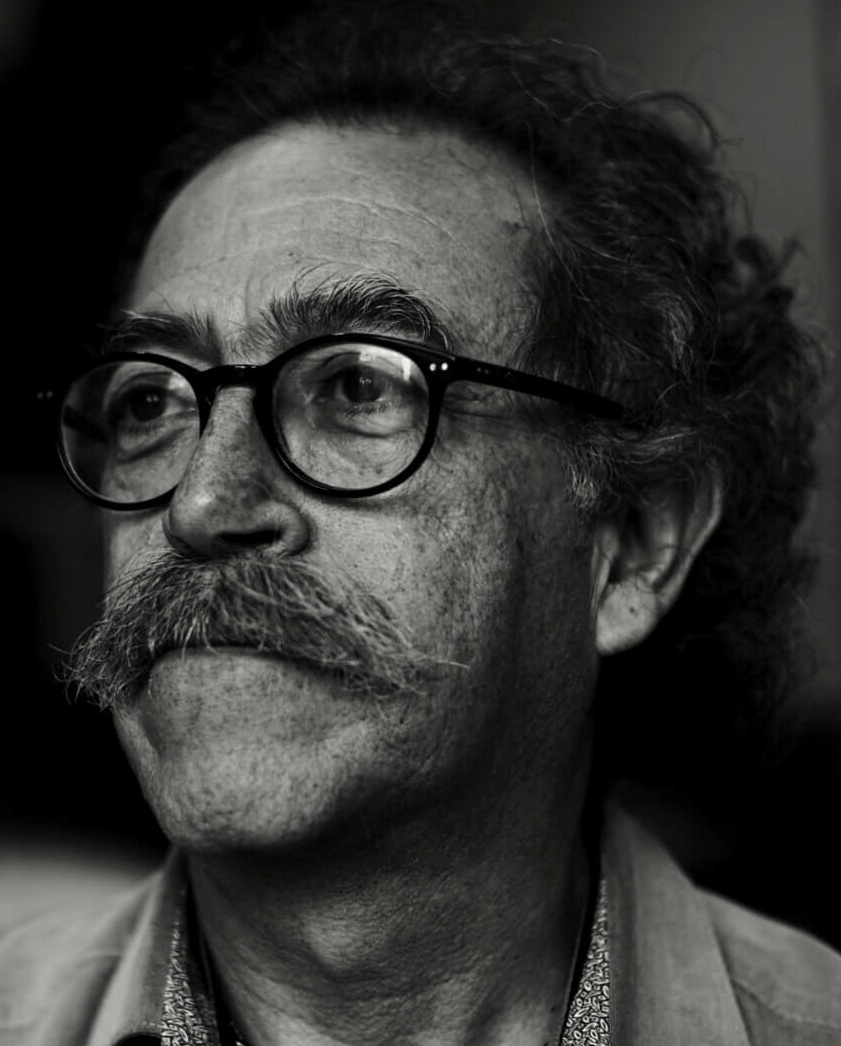}}]{Salvador Cuevas} obtained his BSc on Physics from National Autonomous University of Mexico, an Optical Engineer’s degree from Ecole Supérieure d’Optique/Institut d’Optique Théorique et Appliquée, Orsay, France and a PhD degree from Université Paris-Sud, France. He has been involved on astronomical instrumentation projects for 30 years in Mexico where he has strongly contributed to form an astronomical instrumentation engineering group at the Institute of Astronomy of National Autonomous University of Mexico (UNAM). The last years he has been involved in instrumentation for Gran Telescopio Canarias (GTC). He is author of 20 refereed papers and 88 SPIE technical papers and has been MSc and PhD advisor of over 20 students.
\end{IEEEbiography}

\vfill
\end{document}